\documentclass[aps,prl,twocolumn,groupedaddress,showpacs]{revtex4}
\usepackage{amsfonts}
\usepackage{amssymb}
\usepackage{amsmath}
\usepackage{graphicx}
\usepackage{dcolumn}   
\usepackage{bm}        
\usepackage{flafter}

\begin{document}


\title{Specific Heat of the Ca-Intercalated Graphite Superconductor CaC$_6$}

\date{\today}

\author{J. S. Kim}
\email[Corresponding author:~E-mail~]{js.kim@fkf.mpg.de}
\affiliation{Max-Planck-Institut f\"{u}r Festk\"{o}rperforschung,
Heisenbergstra$\rm \beta$e 1, D-70569 Stuttgart, Germany}
\author{R. K. Kremer}
\affiliation{Max-Planck-Institut f\"{u}r Festk\"{o}rperforschung,
Heisenbergstra$\rm \beta$e 1, D-70569 Stuttgart, Germany}
\author{L. Boeri}
\affiliation{Max-Planck-Institut f\"{u}r Festk\"{o}rperforschung,
Heisenbergstra$\rm \beta$e 1, D-70569 Stuttgart, Germany}
\author{F. S. Razavi}
\affiliation{Department of Physics, Brock University, St.
Catharines, Ontario, L2S 3A1, Canada}

\begin{abstract}

The superconducting state of Ca-intercalated graphite CaC$_6$ has
been investigated by specific heat measurements. The
characteristic anomaly at the superconducting transition ($T_c$ =
11.4 K) indicates clearly the bulk nature of the
superconductivity. The temperature and magnetic field dependence
of the electronic specific heat are consistent with a fully-gapped
superconducting order parameter. The estimated electron-phonon
coupling constant is $\lambda$ = 0.60 - 0.74 suggesting that the
relatively high $T_c$ of CaC$_6$ can be explained within the
weak-coupling BCS approach.

\end{abstract}
\smallskip


\maketitle

The recent discovery of superconductivity in Ca- and
Yb-intercalated graphite has refocussed considerable interest onto
graphite intercalated compounds (GICs)
\cite{YbC6:weller:syn,CaC6:emery:syn}. The superconducting
transition temperature for Ca- and Yb-intercalated graphite is
$T_c$ $\approx$ 11.5 K and 6.5 K, respectively, significantly
higher than the alkali-metal intercalated graphite phases studied
in the 1980's \cite{GIC:dresselhaus:review}. Similar to MgB$_2$
where the hexagonal B sheets are intercalated with Mg, in the GICs
the Ca or Yb atoms are sandwiched by the honeycomb graphene
layers. The intercalated metal ions act as donors and transfer
charge into the host graphene layers, resulting in partially
filled graphene $\pi$-bands. In contrast to MgB$_2$ with a strong
electron-phonon coupling of the B $\sigma$-bands, the coupling
strength of the graphite $\pi$-bands to in-plane phonon modes is
expected to be small. Also out-of-plane phonon modes will not
couple to the $\pi$-band because of the antisymmetric character of
the $\pi$-orbitals. Thus the superconducting mechanism for the
GICs could be rather different than that of MgB$_2$.

The superconducting mechanism, now under debate, relies on the
other partially filled bands: the so-called interlayer bands. From
the detailed comparison between the superconducting and
non-superconducting GICs, Cs$\rm\acute{a}$nyi $et$ $al$.\ found
that the filling of the interlayer state is essential for
stabilizing the superconductivity in GICs \cite{GIC:csanyi:band}.
Based on this result and the consideration about the layered
structure of the GICs, they proposed an unconventional electronic
paring mechanism involving excitons \cite{Exciton:allender:theory}
or low-energy acoustic plasmons \cite{Plasmon:bill:theory}. The
latter has been discussed as a possible mechanism for the high
$T_c$ of the intercalated layered metal halide nitrides
\cite{HfNCl:yamanaka:syn}. Soon after, however, Mazin
\cite{GIC:mazin:band} and Calandra $et$ $al$.\
\cite{CaC6:calandra:band} suggested that the interlayer band
originates from the $s$-band of the intercalant, and an ordinary
electron-phonon coupling mechanism involving the intercalant
phonon and the out-of-plane C phonon modes would be sufficient to
explain the relatively high $T_c$ in the GICs. In order to shed
light on this controversy, further experimental studies on the
superconducting properties of the GICs are required.

Specific heat ($C_P$) measurements are known to be a powerful tool
to investigate the superconducting states. Since $C_P$ probes the
quasi-particle excitation across the superconducting gap, its
temperature and magnetic field dependence reflect the nature of
the superconducting state such as gap symmetry, presence of
multi-gaps, and coupling strength between electrons and phonons
(or other bosonic excitations). So far, there has been no $C_P$
study on the superconducting state of the GICs, mostly because the
low $T_c$ was experimentally difficult to reach, and the sample
homogeneity was not sufficient for reliable studies. In this
Letter, we report the temperature and magnetic field dependence of
$C_P$ for high-quality bulk CaC$_6$ samples. In zero-field
measurements we observe a sharp $C_P$ anomaly at $T_c$ $\sim$ 11.4
K. The detailed temperature and magnetic field dependence of $C_P$
clearly show a fully gapped superconducting order parameter, and
fit very well to the weak coupling BCS prediction.

The Ca-intercalated graphite samples were synthesized by reacting
highly oriented pyrolytic graphite (HOPG, Alfa Aesar, spread of
the $c$-axis orientation 0.4(1)$\rm ^o$) with a molten
lithium-calcium alloy at 350$\rm ^o$ for several weeks ($cf.$
Ref.\ \cite{CaC6:emery:syn}). Since the metal alloy and the
intercalated sample are air sensitive all handling was done in
purified argon atmospheres. X-ray diffraction patterns (not shown)
dominantly show reflections of CaC$_6$ (the distance between
graphene layers, $d$ = 4.50 $\rm \AA$), and contained no
reflections related to non-intercalated graphite. Weak additional
reflections \cite{note:xrd} possibly due to Li/Ca-intercalated
graphite phases \cite{LiCaC:pruvost:syn} were observed. The
contribution of the impurity phase is estimated to be less than
5$\%$ to the total heat capacity. The magnetic susceptibilities
were determined in a MPMS SQUID magnetometer (Quantum Design). The
superconducting transition temperature of our samples is $T_c$ =
11.40(5) K (see the inset of Fig. 1), consistent with previous
reports \cite{CaC6:emery:syn}. The transition width determined as
the temperature difference between 10$\%$ and 90$\%$ diamagnetic
shielding is only 0.1 K, smaller than found in previous reports
\cite{CaC6:emery:syn}. After demagnetization correction, the
volume fraction was estimated to be $\approx$ 100 $\%$. The heat
capacity for a sample of $\approx$ 5 mg was measured using a PPMS
calorimeter (Quantum Design) employing the relaxation method. To
thermally anchor the crystals to the sapphire sample platform and
to orient the samples with respect to the magnetic field ($H$
$\parallel$ $c$), a minute amount of Apiezon N grease was used.
The heat capacity of the platform and the grease, which amounted
to $\approx$ 40-60$\%$ over the whole temperature range, was
determined in a separate run and subtracted from the measurements
for the samples.

\begin{figure}
\includegraphics[height=7.2cm,bb=5 250 580 770]{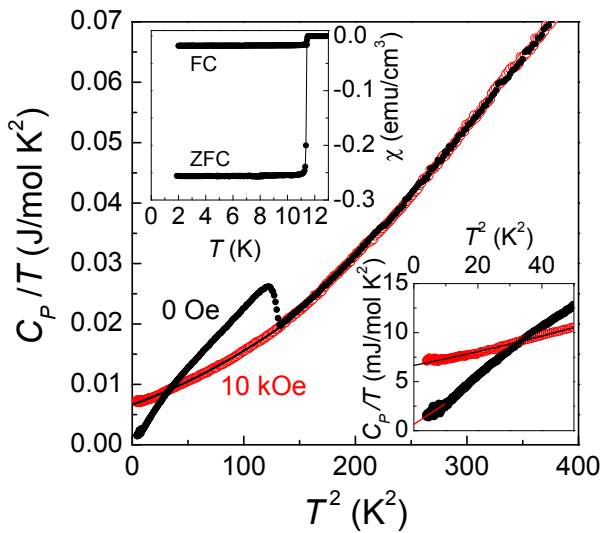}
\caption{(Color online) Temperature dependence of the specific
heat at $H$= 0 and 10 kOe. The solid (black) line through the data
points at $H$ = 10 kOe (also in the right inset) is a fit
described in the text. (Left inset) The temperature dependence of
the susceptibility with $H$ = 6 Oe under zero-field cooled(ZFC)
and field-cooled(FC) modes. (Right inset) The low temperature
behavior of $C_P$($T$). The (red) solid line for $C_P$($T$) at $H$
= 0 is a linear fit for 2 K $<$ $T$ $<$ 3 K.}
\end{figure}

Figure 1 shows the temperature dependence of $C_P$ at $H$ = 0 and
$H$ = 10 kOe applied along the $c$-axis. A sharp anomaly at $T_c$
is resolved indicating the bulk nature of the superconductivity in
our sample. The onset of 11.40 K determined from the specific heat
jump is consistent with the $T_c$ obtained from the susceptibility
measurements. A small offset of $C_P$/$T$ at $H$= 0 ($\gamma_{ns}$
= 0.65 $\pm$ 0.13 mJ/mol K$^2$) and a slight hump of $C_P$/$T$ at
$H$ = 10 kOe were observed below 3 K (see the right inset of Fig.
1), which probably originate from paramagnetic impurities or a
non-superconducting fraction in our sample. $\gamma_{ns}$ is only
$\sim$ 10 $\%$ of the Sommerfeld coefficient $\gamma_N^*$
extracted from the normal state $C_P$($T)$ as discussed below. If
we assume that $\gamma_{ns}$ corresponds to the
non-superconducting part of our sample, the volume fraction of the
superconducting portion can be estimated as
1$-$($\gamma_{ns}$/$\gamma_N^*$) $\approx$ 90$\%$ proving a good
quality of our GIC sample. The upper critical field
$H_{c2}^{\perp}$ along the $c$-axis has been reported to be $\sim$
2.5 kOe \cite{CaC6:emery:syn}, thus the normal state $C_P$($T)$
can be obtained when the superconductivity is completely
suppressed by a magnetic field $H$ = 10 kOe. A clear deviation
from Debye $T^3$ law seen in the normal state $C_P$ can be
attributed to a contribution from the low-lying optical phonon
modes. Recent $ab$-initio calculations also show the presence of
low frequency modes mainly involving vibration of the intercalated
Ca \cite{CaC6:calandra:band}. The normal state $C_P$ can be
described by $C_P$ = $\gamma_N^*$$T$ + $C_{lattice}$($T$) where
$\gamma_N^*$$T$ is the electronic contribution, and
$C_{lattice}$($T$) = $\beta$$T^3$ + $\delta$$T^5$ is the lattice
contribution. The solid line in Fig. 1 is the best fit to the $H$
= 10 kOe data below $T$ $<$ 12 K, yielding the parameters,
$\gamma_N^*$ = 6.66(1) mJ/mol K$^2$, $\beta$ = 65.1(3) $\mu$J/mol
K$^4$, and $\delta$ = 0.256(3) $\mu$J/mol K$^6$. Considering the
offset of $C_P$/$T$($H$ = 0), $\gamma_{ns}$, the Sommerfeld
coefficient for the superconducting part of the sample is
$\gamma_N$ $\approx$ $\gamma_N^*$$-$$\gamma_{ns}$ = 6.01 $\pm$
0.14 mJ/mol K$^2$. The corresponding Debye temperature
$\Theta_D$(0) is 598(3) K, which is much higher than that of pure
graphite ($\Theta_D$(0) $=$ 413 K) and the alkali metal GICs
($e.$$g.$ 235 K for KC$_8$) \cite{GIC:mizutani:Cp}, but comparable
with $\Theta_D$(0) $=$ 590 - 710 K for LiC$_6$
\cite{LiC6:Delhaes:Cp}.

\begin{figure}
\includegraphics[height=7 cm,bb=25 280 530 780]{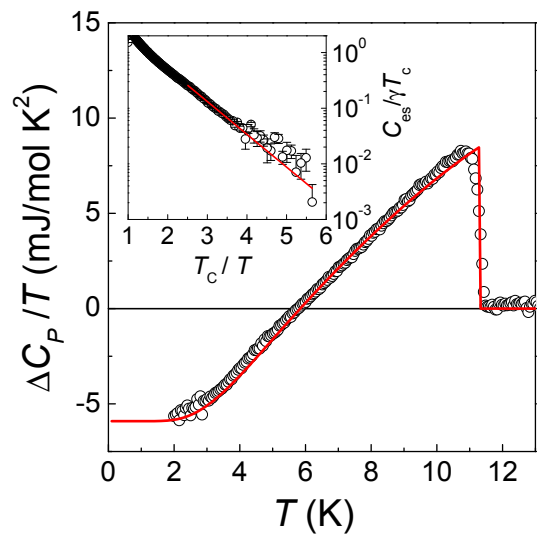}
\caption{(Color online) Temperature dependence of
$\Delta$$C_P$/$T$ = $C_P$($H$= 0)/$T$ - $C_P$($H$ = 10 kOe)/$T$.
The (red) solid line is the best fit according to the
$\alpha$-model assuming an isotropic $s$-wave BCS gap as described
in the text. The electronic contribution of the specific heat
$C_{es}$ is plotted on a logarithmic scale against $T_c$/$T$ in
the inset. The (red) straight line in the inset is the exponential
fit for 2.5 $<$ $T_c$/$T$ $<$ 5.5.}
\end{figure}

The specific heat difference $\Delta C_P$ between the normal and
superconducting state is shown in Fig. 2. The solid line is the
theoretical fit based on the `$\alpha$-model' assuming an
isotropic $s$-wave BCS gap $\Delta$($T$) scaled by the adjustable
parameter, $\alpha$ = $\Delta$(0)/$k_B$$T_c$
\cite{Cp:padamsee:alpha}. For the weak coupling limit $\alpha$ is
1.76. The detailed temperature dependence of $\Delta C_P$/$T$ was
fitted by three adjustable parameters: $\alpha$, Sommerfeld
coefficient $\gamma$, and $T_c$. The data are very well reproduced
by $\alpha$ = 1.776, $\gamma$ = 5.91 mJ/mol K$^2$ and $T_c$ =
11.30 K. The normalized specific heat jump,
$\Delta$$C_P$/$\gamma$$T_c$ is 1.432, which is close to the weak
limit BCS value, 1.426. The normalized electronic specific heat,
$C_{es}$/$\gamma$$T_c$ in the superconducting state is shown in
the inset. The solid line is an exponential fit to the data for
2.5 $\leq$ $T_c$/$T$ $\leq$ 5.6 using the form
$C_{es}$/$\gamma$$T_c$ $\propto$ $\exp$(-0.82$\alpha$$T_c$/$T$)
with $\alpha$ =1.65(1). $C_{es}$ $exponentially$ $vanishes$ for
$T$ $\rightarrow$ 0 K, clearly manifesting the absence of gap
nodes in the superconducting state. The $\alpha$ value is somewhat
lower than that from the $\alpha$-model fit in which the ratio
$\Delta$(0)/$k_B$$T_c$ is mostly determined by the shape of the
$C_P$ jump near $T_c$. However the discrepancy is less than
10$\%$, in contrast to MgB$_2$ where the discrepancy is more than
60$\%$ due to the presence of two gaps \cite{MgB2:bouquet:Cp}.

\begin{figure}
\includegraphics[height=7.5cm,bb=0 255 585 770]{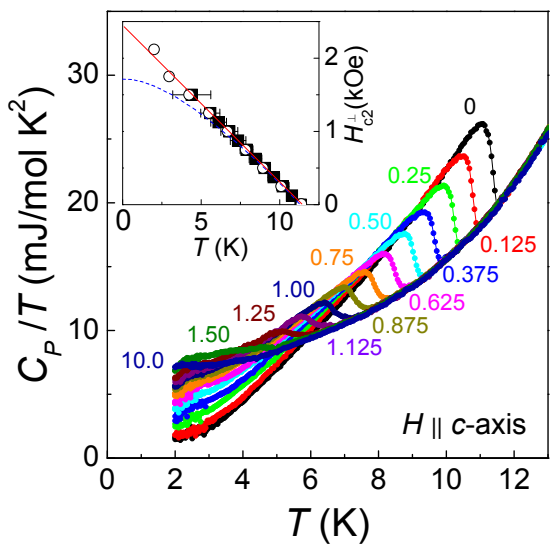}
\caption{(Color online) Magnetic field dependence of the heat
capacity for CaC$_6$. The numbers next to the data correspond to
the applied magnetic field (kOe) along the $c$-axis. The inset
shows the $H_{c2}^{\perp}$($T$) for $H$ $\parallel$ $c$ estimated
from specific heat ($\blacksquare$) and susceptibility
($\bigcirc$). The (blue) dashed line demonstrates the WHH
prediction \cite{Hc2:WHH}, and the (red) solid line is a linear
fit for the low magnetic field data ($H$ $\leq$ 1 kOe).}
\end{figure}

Figure 3 shows the temperature dependence of $C_P$/$T$ measured
with different magnetic fields. Superconductivity is gradually
suppressed, and the quasiparticle contribution increases with
magnetic field. $T_c$($H$) determined by the mid-point over the
$C_P$ anomaly is plotted in the inset with the error bar
corresponding to the transition width. The $T_c$($H$) obtained
from the susceptibility measurements(not shown) are in good
agreement with those found from $C_P$($T$). For comparison, we
also plot the predicted curve based on the
Werthamer-Helfand-Hohenberg (WHH) theory \cite{Hc2:WHH}.
$H_{c2}^{\perp}$($T$) shows a linear dependence down to $T$/$T_c$
$\approx$ 0.2 and a clear deviation to the WHH prediction, which
deserves further investigation. The best fit for low fields
results in ($d H_{c2}^{\perp}$/$d T$)$_{T_c}$ = 219.1(8) Oe/K and
the linear extrapolation of $H_{c2}^{\perp}$($T$) to $T$ = 0 K
yields $H_{c2}^{\perp}$(0) = 2.48 kOe, which corresponds to an
$ab$-plane coherence length, $\xi_{ab}$(0) $\approx$ 360 ${\rm
\AA}$. From $H_{c2}^{\parallel}$(0) $\approx$ 7 kOe along the
$ab$-plane \cite{CaC6:emery:syn}, the $c$-axis coherence length,
$\xi _{c}$(0) $\sim$ 130 $\rm \AA$ ($\gg$ $d$ = 4.50 $\rm \AA$) is
obtained. This result indicates the 3-dimensional nature of
superconductivity in CaC$_6$ \cite{CaC6:emery:syn}.

\begin{figure}
\includegraphics[height=7.0cm,bb=10 330 580 775]{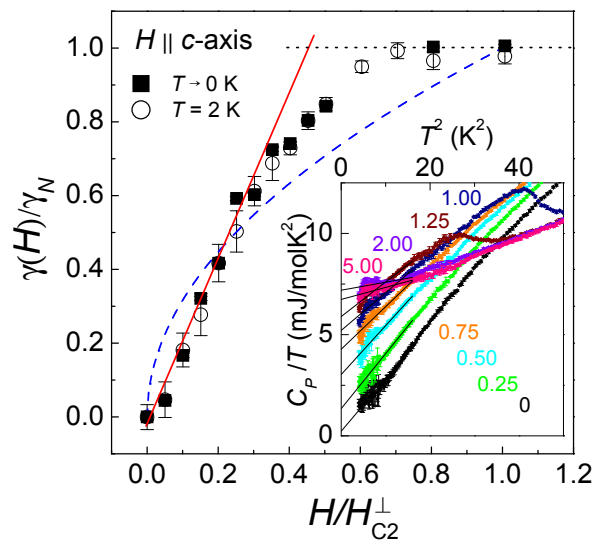}
\caption{(Color online) Magnetic field dependence of $\gamma$($H$)
as a function of $H$/$H_{c2}^{\perp}$. $\gamma (H)$ for $T$
$\rightarrow$ 0 K ($\blacksquare$) and at $T$ = 2 K ($\bigcirc$)
is normalized by $\gamma_N$ = $\gamma$(10 kOe). The (red) solid
and (blue) dashed lines correspond to $\gamma(H)$ $\propto$ $H$
and $\gamma(H)$ $\propto$ $H^{1/2}$, respectively. The (black)
dotted line indicates $\gamma(H)$/$\gamma_N$ = 1. (Inset) Low
temperature $C_P$ data at various magnetic fields. The (black)
solid lines are fits to the data for 2 K $\leq$ $T$ $\leq$ 4 K.
The numbers next to the data correspond to the magnetic fields
(kOe).}
\end{figure}

The magnetic field dependence of the Sommerfeld coefficient
$\gamma$($H$) provides information about the superconducting gap
symmetry. For a gapped superconductor, the quasiparticle
excitations are expected to be confined in the vortex cores, and
$\gamma$($H$) is proportional to the density of vortices resulting
in $\gamma (H)$ $\propto$ $H$. In contrast, when the
superconducting gap is highly anisotropic or has a gap node, the
delocalized quasiparticles near the gap minima cause a nonlinear
magnetic field dependence such as $\gamma$($H$) $\propto$
$H^{1/2}$ \cite{Cp:volovik:gammaH}. The normalized $\gamma$($H$)
for CaC$_6$ is displayed in Fig. 4. $\gamma (H)$ for $T$
$\rightarrow$ 0 K is determined from the linear fit for 2 K $\leq$
$T$ $\leq$ 4 K (see the inset of Fig. 4) after subtracting the
offset of $\gamma_{ns}$($H$ = 0). As a consistency check, we also
plot $C_P(T,H)$/$T$ ($\approx$ $C_{es}$($T$,0)/$T$ +
$\gamma$($T$,$H$) + $C_{lattice}$/$T$) at the lowest temperature
in the present measurements ($T$ = 2 K) after subtraction of
$C_P$($H$ = 0)/$T$($\approx$ $C_{es}$($T$,0)/$T$ +
$C_{lattice}$/$T$). Both curves are normalized with $\gamma$($H$ =
10 kOe).

As shown in Fig. 4, $\gamma$($H$) increases linearly with $H$ for
low fields up to $H$ $\approx$ 0.3 $H_{c2}$. This behavior is
strongly adverse to $\gamma$($H$) $\propto$ $H^{1/2}$ expected for
nodal superconductivity and also in contrast to two-band
superconductivity \cite{MgB2:yang:Cp}. Recent calculations found
that even for an isotropic-gapped type-II superconductor,
$\gamma(B)$ $\propto$ $B$ behavior persists only up to a certain
crossover field, $B^*$ \cite{Cp:nakai:gammaH}. For the isotropic
superconducting gap, $B^*$ is expected to be $\approx$
0.32$B_{c2}$ which is reduced as the degree of the anisotropy for
the superconducting gap increases. Since $C_P(H)$ measurements
were done in the field-cooled mode, we can approximately estimate
the ratio between the crossover field and the upper critical field
$B^*$/$B_{c2}$ $\approx$ $H^*$/$H_{c2}$ $\sim$ 0.3, which is quite
close to the isotropic limit. Thus the magnetic field dependence
of $\gamma$($H$) consistently supports the notion of a
fully-gapped and almost isotropic superconducting order parameter.

Finally, we estimate the electron-phonon coupling strength,
$\lambda$ for CaC$_6$. At first, $\lambda$ can be estimated from a
comparison between $\gamma_N$ and the density of states at the
Fermi level $E_F$, $N(0)$ using the equation $\gamma_N$ =
(2$\pi^2$$k_B^2$/3)$N$(0)(1+$\lambda$). If we take $N(0)$ = 1.50
states/eV$\cdot$cell from recent calculations
\cite{CaC6:calandra:band}, we obtain $\lambda$ $\approx$ 0.66 -
0.74 using our $\gamma_N$ $=$ 6.01 $\pm$ 0.14 mJ/mol K$^2$.
Alternatively, we can also estimate $\lambda$ based on the
McMillan formula \cite{Tc:mcmillian:theory},
\begin{equation}
T_c =\frac{\Theta_{D}}{1.45} \exp \left[
\frac{-1.04(1+\lambda)}{\lambda-(1+0.62\lambda)\mu^*} \right].
\end{equation}
Here, $\mu^*$ is the Coulomb pseudopotential. Taking $T_c$ = 11.4
K and  $\Theta_D$ = 598(3) K from the normal state $C_P$, we
obtain $\lambda$ = 0.60 - 0.71 with $\mu^*$ ranging from 0.10 to
0.15. All these results are in good agreement with each other, and
imply that CaC$_6$ obviously belongs to the weak coupling regime.
Recent calculations \cite{CaC6:calandra:band} also predicted
$\lambda$ to be $\approx$ 0.83, in accordance with the measured
$\lambda$. Consistently, the superconducting gap ratio,
2$\Delta$(0)/$k_B$$T_C$ = 3.3 - 3.6 is also close to the BCS value
3.52.

In conclusion, we have reported the first specific heat
measurements for the superconducting and normal state of
high-quality bulk CaC$_6$ samples. The specific heat anomaly at
$T_c$ is clearly resolved indicating the bulk nature of the
superconductivity. Both the temperature and magnetic field
dependence of $C_P$ strongly suggests a fully-gapped
superconducting order parameter. The estimated electron-phonon
coupling constant $\lambda$ is 0.60 - 0.74, in a good agreement
with recent calculations. These results suggest that CaC$_6$ is a
fully-gapped, weakly-coupled, phonon-mediated superconductor
without essential contributions from alternative paring
mechanisms. We cannot rule out the possibility of a
superconducting gap anisotropy or a multi-gap scenario in this
system \cite{note:Cp}, however, if any, the difference between the
largest and smallest gaps will be rather small. Very recently,
Lamura $et$ $al$.\cite{CaC6:lamura:penetration} reported the
in-plain penetration depth, $\lambda_{ab}$ for the bulk CaC$_6$
sample. A clear exponential temperature dependence of
$\lambda_{ab}$ indicates a fully-gapped superconductivity in
CaC$_6$, which is consistent with our $C_P$ study.

\acknowledgments

The authors acknowledge D.\ Guerard and A.\ Simon for sharing
ideas about intercalation chemistry. We also thank E. Br\"{u}cher,
G. Siegle, and C. Hoch for experimental supports and W. Schnelle,
B. Mitrovi$\rm \acute{c}$, O. V. Dolgov, and O. K. Andersen for
useful discussions.


\begin{thebibliography}{}
\bibitem{note:xrd} The intensity ratio of the reflections from the impurity phase and the CaC$_6$ phase is $<$ 1$\%$.
\bibitem{note:Cp} The anisotropic or
two- gap nature might be smeared out because a certain level of
disorder is inevitable, $e.$ $g.$ the intercalated Ca
distribution.
\end{thebibliography}
\end{document}